\documentclass{iau} 
\usepackage{graphicx}

\newcommand{\apj}{\textit{ApJ}}
\newcommand{\apjs}{\textit{ApJS}}
\newcommand{\aap}{\textit{A\&A}}
\newcommand{\araa}{\textit{ARA\&A}}
\newcommand{\mnras}{\textit{MNRAS}}

\title[Studying the SN-SNR connection through multi-D modeling] %% give here short title %%
{Bridging the gap between supernovae and their remnants through
multi-dimensional hydrodynamic modeling}

\author[S. Orlando, M. Miceli, and O. Petruk]   %% give here short author list %%
{S. Orlando$^1$, M. Miceli$^{2,1}$, \and O. Petruk$^{1,3}$}

\affiliation{$^1$INAF-Osservatorio Astronomico di Palermo, Piazza
del Parlamento 1, 90134 Palermo, Italy \\
$^2$Dip. di Fisica e Chimica, Univ. di Palermo, Piazza del Parlamento
1, 90134 Palermo, Italy\\
$^3$Inst. Appl. Probl. in Mechanics and Mathematics,
Naukova Street, 3-b Lviv 79060, Ukraine}

\pubyear{2015}
\volume{xxx}  %% insert here IAU Symposium No.
\jname{Title of your IAU Symposium}
\editors{A.C. Editor, B.D. Editor \& C.E. Editor, eds.}
\begin{document}

\maketitle

\begin{abstract}

Supernova remnants (SNRs) are diffuse extended sources characterized
by a complex morphology and a non-uniform distribution of ejecta.
Such a morphology reflects pristine structures and features of the
progenitor supernova (SN) and the early interaction of the SN blast
wave with the inhomogeneous circumstellar medium (CSM). Deciphering
the observations of SNRs might open the possibility to investigate
the physical properties of both the interacting ejecta and the
shocked CSM. This requires accurate numerical models which describe
the evolution from the SN explosion to the remnant development and
which connect the emission properties of the remnants to the
progenitor SNe. Here we show how multi-dimensional SN-SNR hydrodynamic
models have been very effective in deciphering observations of SNR
Cassiopeia A and SN 1987A, thus unveiling the structure of ejecta
in the immediate aftermath of the SN explosion and constraining the
3D pre-supernova structure and geometry of the environment surrounding
the progenitor SN.
\end{abstract}

\section{Introduction}

Supernova remnants (SNRs) are probably among the most beautiful
objects which is possible to observe in the sky. Their beauty is
mostly due to the complex morphology and highly non-uniform
distribution of ejecta which characterize most of them. In fact the
propagation of ejecta drives shocks back and forth in the interstellar
medium (ISM) and through ejecta themselves, leading to features and
complex structures virtually at all spatial scales. The physics
governing SNRs is studied thanks to observations at all wavelength
bands, from radio, to optical, to X- and $\gamma$-rays. Multi-wavelength
observations allow us to study both the thermal and non-thermal
emission arising from SNRs. The physical and chemical properties
of the emitting plasma are inferred through the analysis of both
low and high resolution spectra. The observations allow us to
investigate the morphology and the projected distributions of
properties in the remnants and to study the dynamics, energetics,
and evolution of SNRs. Besides the possibility to investigate the
physical processes at work in SNRs, there is a large consensus that
the observations encode information also on the progenitor star and
the SN explosion from which the remnant originates. Obvious questions
are: What information might we hope to obtain by deciphering the
observations? How can we decipher the observations?

In fact, the morphological properties of SNRs may reflect pristine
structures and features of the progenitor SN explosion and the
physical and chemical properties of the progenitor SN. Thus,
investigating the intimate link that exists between the morphological
properties of a SNR and the complex phases in the SN explosion may
allow us to trace back the structure and chemical composition of
SN ejecta, and the dynamics and energetics of the SN explosion. An
example is the SNR Cassiopeia A (Cas\,A) whose complex morphology
is mainly due to a highly inhomogeneous distribution of ejecta
determined most likely in the immediate aftermath of the SN explosion.
In the case of young remnants, the morphological properties of SNRs
may also reflect the early interaction of the SN blast with the
inhomogeneous circumstellar medium (CSM) formed during the latest
stages of the stellar progenitor’s evolution. In this case, analyzing
the morphology of a SNR may help us to probe the structure and
geometry of the CSM immediately surrounding the SN, providing
important clues on the final stages of stellar evolution. An example
is the remnant of SN 1987A whose morphology reflects the interaction of the
blast wave with the inhomogeneous CSM consisting mainly of an H~II
region and a dense equatorial ring.

All the above considerations point out that linking SNR morphology
to their SN progenitors is a necessary step: 1) to probe the physics
of SN engines by providing insight into the asymmetries that occur
during the SN explosion, and 2) to investigate the final stages of
stellar evolution by unveiling the structure of the medium immediately
surrounding the progenitor star. This step is now essential to open
new exploring windows on SN and SNR issues especially in view of the
spatially resolved high-resolution spectroscopy capability of the
forthcoming Athena satellite (\cite{2013arXiv1306.2307N}). In the
following, we review our recent achievements in this field.

\section{The strategy to link supernovae to supernova remnants}
\label{sec:strategy}

How to link SNe to SNRs? In general existing hydrodynamic and
magnetohydrodynamic models have described either the SN evolution
or the expansion of the remnant. Former models describe the complex
phases of the SN evolution and they do not follow its subsequent
transition to the phase of SNR and the interaction between the SN
blast wave and the CSM. Conversely, models of SNRs describe the
interaction of the remnant with the environment; however they usually
assume an initial parameterized ejecta profile, typically few years
after the SN event, and leave out an accurate description of the
ejecta structure and chemical stratification soon after the SN
explosion. Unfortunately this sharp distinction between these two
classes of models and the poor communication between the community
studying the physics of SNe and that of SNRs have prevented so far
to disentangle the effects of the initial conditions (namely the
SN event) from those of the boundary conditions (namely the interaction
of the SN blast wave with the environment).

In fact there are several criticalities in describing the whole
phenomenon from the on-set of the supernova to the full remnant
development. One serious problem is to catch the very different
scales in time and space of SNe and SNRs. This has hampered so far
to study the SN-SNR connection in detail. The core collapse of a
massive star occurs on time scales of the order of a second and on
dimensions of the order of 10000 km (e.g. \cite{2016ARNPS..66..341J}).
At the other end, a young SNR evolves on time scales of the order
of hundreds of years (if not thousands of years in the case of more
evolved remnants) and can be characterized by dimensions of the
order of parsecs. In addition the phenomenon is inherently
three-dimensional (3D), thus requiring for the modeling a huge
amount of numerical resources. This is the reason why the rare
attempts to study the SN-SNR connection have been made, in the past, by
adopting a one-dimensional (1D) approach (e.g. \cite{2008ApJ...680.1149B,
2014ApJ...785L..27Y, 2015ApJ...803..101P}). However, the 3D nature
of SN explosions is well known, being these characterized by strong
asymmetries (jets, knots, clumps; e.g. \cite{1993ApJ...419..824L,
2002ApJ...579..671W, 2006A&A...453..661K, 2008ARA&A..46..433W,
2016ARNPS..66..341J}); hydrodynamic instabilities easily develop
after the core-collapse, producing dense fingers of ejecta (e.g.
\cite{2016ARNPS..66..341J}); and turbulence mixing of ejecta is
most likely present (e.g.  \cite{2016ARNPS..66..341J}). During the
remnant expansion, the clumpy structure of the ejecta may induce
complex interactions among the clumps and shrapnels of ejecta,
enhancing the development of hydrodynamic instabilities, the mixing
of initially chemically homogeneous layers, and some overturning
of the ejecta due to hydrodynamic instabilities developing during
clumps interaction (e.g. \cite{2012ApJ...749..156O, 2013MNRAS.430.2864M}).
In addition the ambient medium (CSM or ISM) through which the SN
blast wave expands is, in general, highly inhomogeneous and
characterized, for instance, by nebulae, rings, clouds, or gradients
of density.  The ambient magnetic field is, in general, non-uniform;
even in the case of a uniform magnetic field the obliquity angle
between the magnetic field and the shock normal varies along the
remnant outline (e.g. \cite{2009MNRAS.393.1034P}). Hydrodynamic instabilities
develop at the contact discontinuity between the stellar debris and
the shocked ambient medium and at the border of shocked CSM/ISM
inhomogeneities (e.g. \cite{2005A&A...444..505O, 2008ApJ...678..274O}).

\begin{figure}[t]
\begin{center}
 \includegraphics[width=13.cm]{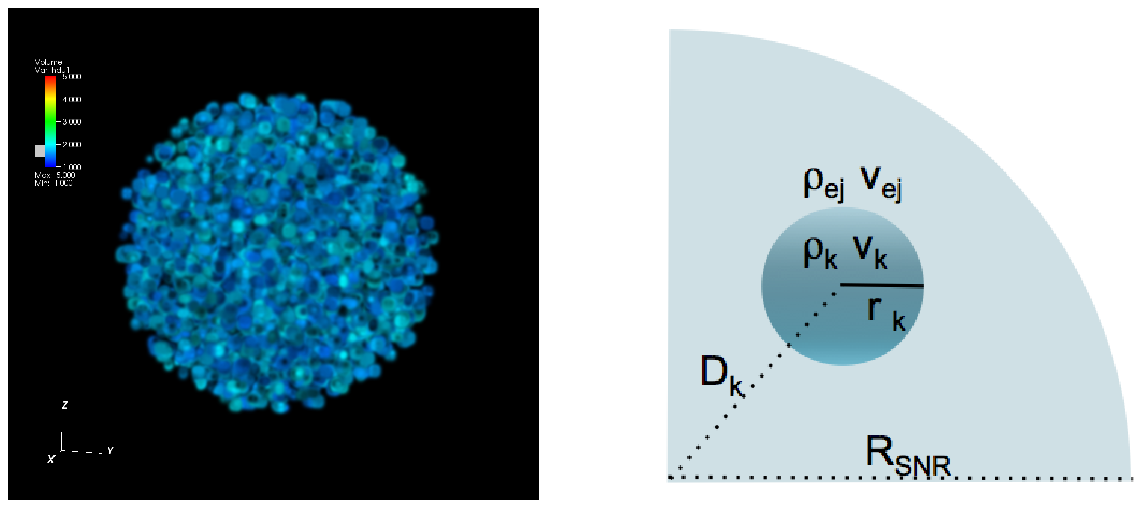}
 \caption{Initial 3D post-explosion structure of the ejecta. The
 panel on the left shows the small-scale clumping prescribed as in
 Orlando et al. (2012). The panel on the right shows a schematic
 view of the large-scale anisotropies described as overdense spherical
 knots characterized by radius, $r_{k}$, distance from the center
 of explosion, $D_k$, and density and velocity contrasts with
 respect to the surrounding ejecta, $\rho_k/\rho_{ej}$ and $v_k/v_{ej}$
 respectively.}
   \label{fig_knots}
\end{center}
\end{figure}

In the attempt to study the SN-SNR connection, recently we have
devised a strategy which is based on the coupling between a 1D model
of core-collapse SN and a 3D model of SNR (\cite{2015ApJ...810..168O,
2016ApJ...822...22O}). The SN model describes the evolution of a
core-collapse SN from the breakout of the shock wave at the stellar
surface to the so-called nebular phase. The model is implemented
through a relativistic, radiation-hydrodynamics Lagrangian code
(\cite{2011ApJ...741...41P}) which includes a full general relativistic
treatment and the radiative transfer coupled to relativistic
hydrodynamics at all regimes. The model includes also the gravitational
effects of the central compact remnant on the evolution of the
ejecta (fall-back of material onto the compact remnant; amount of
ejected $^{56}$Ni). The 1D SN model provides the initial radial
distribution of ejecta for the 3D SNR model. The latter describes
the transition from the phase of SN to that of SNR and the interaction
of the remnant with the (inhomogeneous) ambient medium. The 3D
SNR model is implemented with the FLASH code (\cite{for00})
and includes the effects of radiative losses from optically thin
plasma, the deviations from equilibrium of ionization, the back
reaction of accelerated cosmic rays, and the deviations from
electron-proton temperature equilibration. The evolution of
an appropriate post-explosion isotopic composition of the ejecta
is accounted by performing multi-species simulations. The model
describes the post-explosion structure of the ejecta by small-scale
clumping of material and larger-scale anisotropies (see
Fig.~\ref{fig_knots}). The evolution of different plasma components
(e.g. ejecta, inhomogeneities of CSM, etc.) is traced by considering
passive tracers associated with each of them. From the model results,
the thermal emission is synthesized in different bands. In this
way, the model results can be compared with the observations to
obtain a feedback on the SN-SNR models.

A major challenge in our approach is capturing the enormous range
in time and space scales. The initial condition for the 3D simulations
of the SNR is typically 1 day after the SN event. At that time the
size of the ejecta distribution is of the order of tens of AU. Then
the remnant evolution is followed for hundreds of years. At the
final time, the size of the remnant can be of several pc. To catch
this huge range of time and space scales, our approach exploits the
adaptive mesh refinement capabilities of the FLASH code. About 20
nested levels of adaptive mesh refinement are typically used to
reach an effective resolution of about 0.2~AU ($\approx 3\times
10^{12}$~cm). This guarantees to have at least 100 computational
cells per remnant radius during the whole evolution.

As first applications of our SN-SNR model, we have considered two
study cases which are, in some way, complementary each other. The
first is the SNR Cas\,A whose morphology is believed to reflect
pristine structures and features developed in the immediate aftermath
of the SN explosion. The second case is that of SN 1987A which is
characterized by a very complex morphology which reflects the
interaction of the SN blast wave with the strongly inhomogeneous
CSM. These simulations required about 7.7 millions of CPU hours on
the MareNostrum III cluster hosted at the Barcelona Supercomputing
Center (Spain) and about 4 millions of CPU hours on the
FERMI cluster hosted at CINECA (Italy).

\section{Effects of supernova anisotropies: The case of SNR Cassiopeia A}

The observations suggest that the morphology and expansion rate of
Cas\,A are consistent with a remnant expanding through the wind of
the progenitor red supergiant (\cite{2014ApJ...789....7L}).  Since
the structure of the wind is expected to be spherically symmetric
with the gas density proportional to $r^{-2}$ (where $r$ is the
radial distance from the center of explosion), the complex morphology
of the remnant and the highly non-uniform distribution of ejecta
should originate from anisotropies developed soon after the SN
explosion. For this reason, Cas\,A is considered a very attractive
laboratory to bridge the gap between SNe and their remnants. In
addition Cas\,A is one of the best studied SNR and its 3D structure
has been characterized in good detail thanks to observations in the
IR, optical, and X-ray bands (e.g. \cite{2010ApJ...725.2038D,
2012ApJ...746..130H, 2013ApJ...772..134M}). In particular, the
observations show that the ejecta distribution is characterized by
three extended Fe-rich regions and two Si-rich jets. In addition,
rings of Si-, S-, and O-rich ejecta surround the Fe-rich regions
(\cite{2013ApJ...772..134M}).

Recently we have investigated the origin of the most pronounced
features observed in the actual distribution of ejecta in Cas\,A
following the approach outlined in Sect.~\ref{sec:strategy}
(\cite{2016ApJ...822...22O}). First the 1D SN model was used to
provide the initial radial profile of ejecta for the 3D SNR model,
about 1 day after the SN event. Then the 3D post-explosion structure
of the ejecta was described by small-scale clumping of material and
larger-scale anisotropies (\cite{2006A&A...453..661K,
2008ARA&A..46..433W}). The small-scale clumping was prescribed as
in Orlando et al. (2012, see left panel in Fig.~\ref{fig_knots}).
The large-scale anisotropies were described as overdense spherical
knots characterized by the following parameters: radius of the knot,
distance of the knot from the center of explosion, density and
velocity contrasts of the knot with respect to the surrounding
ejecta (see right panel in Fig.~\ref{fig_knots}). By exploring the
model parameters space, first we have derived the most likely mass
of ejecta ($M_{\rm ej}\approx 4M_{\odot}$) and explosion energy
($E_{\rm SN}\approx 2.3\times 10^{51}$~erg) which are those able
to reproduce simultaneously the radii and velocities of the forward
and reverse shocks as observed at the current time
(\cite{2009A&A...503..495V}) and the density of the shocked red
supergiant wind inferred from observations (\cite{2014ApJ...789....7L}).
Then we have constrained the physical parameters (namely size,
density, velocity, and position in the original onion-skin
nucleosynthetic layering) characterizing the anisotropies developed
soon after the SN explosion and responsible for the observed masses
and distributions of Fe and Si/S in Cas\,A.

\begin{figure}[t]
\begin{center}
 \includegraphics[width=13cm]{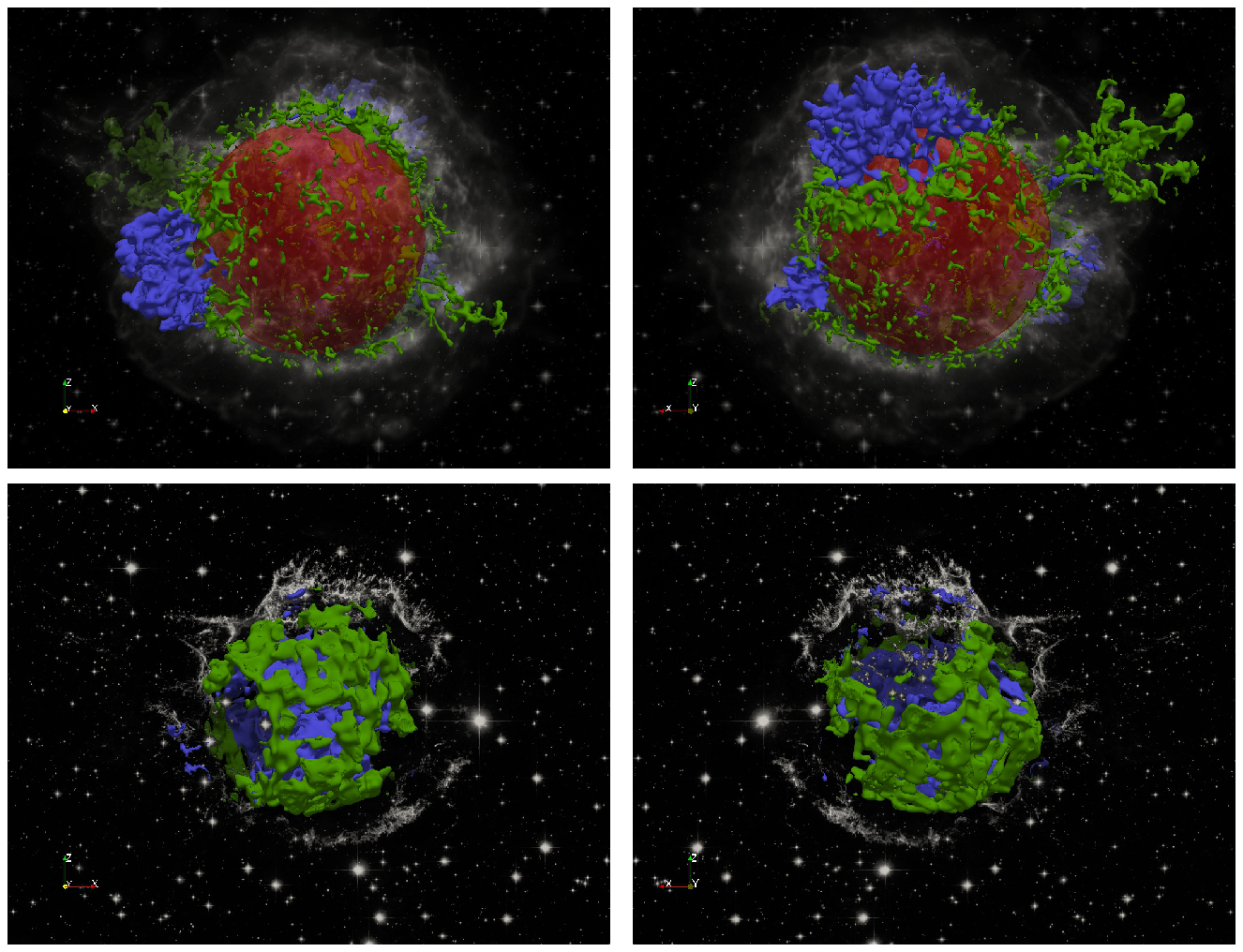} 
 \caption{3D spatial distribution of Cas\,A shocked (upper panels)
 and unshocked (lower panels) Fe (blue) and Si/S (green) rich ejecta
 derived from the best-fit model (adapted from Orlando et al. 2016).
 The figure shows the 3D distributions assuming the perspective in
 the plane of the sky (on the left) or from behind Cas\,A (on the
 right). The transparent red sphere marks the fiducial reverse
 shock. The transparent image in the upper panels is a Chandra
 observation showing the hot shocked plasma in the wavelength band
 $[0.3, 10]$ keV (retrieved from www.nasa.gov); the transparent
 image in the lower panels is a composite Hubble Space Telescope
 (HST) image sensitive to emission in cold O and S lines (retrieved
 from www.spacetelescope.org).}
   \label{fig_casa}
\end{center} \end{figure}

Figure~\ref{fig_casa} shows the distribution of both shocked and
unshocked Fe-rich and Si/S-rich ejecta at the age of Cas A. We have
found that post-explosion anisotropies (pistons) reproduce the
observed distributions and masses of Fe and Si/S if their total
mass was $\approx 0.25\,M_{\odot}$ (corresponding to $\approx 5$\%
of the total mass of ejecta) and their kinetic energy $\approx 1.5\times
10^{50}$~erg (corresponding to $\approx 7$\% of the total energy of
explosion). The model has shown that the pistons produce a spatial
inversion of ejecta layers at the epoch of Cas A, leading to the
Si/S-rich ejecta physically interior to the Fe-rich ejecta, in nice
agreement with the observations of Cas\,A (\cite{2010ApJ...725.2038D,
2013ApJ...772..134M}). In addition the model explains the origin
of the rings of Si/S-rich material encircling the Fe-rich regions
(\cite{2013ApJ...772..134M}): they form at the intersection between
the reverse shock and the material accumulated around the pistons
during their propagation. As for the unshocked ejecta, the model
has shown that the distribution of Si/S is characterized by large
cavities corresponding to the directions of propagation of the
pistons/jets. This may explain why the cavities observed in
near-IR observations are physically connected to the bright rings
in the main-shell (\cite{2015Sci...347..526M}). The model, therefore,
supports the idea that the bulk of asymmetries observed in Cas\,A
are intrinsic to the explosion.

It is interesting to note that, very recently, 3D simulations of a
neutrino-driven SN explosion were able to reproduce some basic
properties of Cas\,A (\cite{2016arXiv161005643W}). The initial
conditions of this model are given immediately after the core bounce
and the simulations follow the evolution until $\approx 1$~day after
the explosion, namely roughly the time when we have prescribed the
post-explosion anisotropies in our 3D SNR simulations
(\cite{2016ApJ...822...22O}). In one of their simulations,
Wongwathanarat et al. (2016) have found three pronounced Fe-rich
fingers that may correspond to the extended Fe-rich regions observed
in Cas\,A. Interestingly their physical characteristics seem to
be similar to those of initial pistons prescribed in our SN-SNR
model, 1\,day after the SN explosion.

\section{Effects of inhomogeneous CSM: The case of SN 1987A}

The effects of an inhomogeneous CSM on the morphology of a SNR can
be investigated in SN\,1987A. In fact, in this case, the surrounding
medium (the nebula) is highly inhomogeneous as shown by optical
images soon after the outburst. The nebula consists mainly of a
dense central equatorial ring surrounded by an extended H\,II region.
Two other rings lying in planes almost parallel above and below the
equatorial one are also observed. This complex structure of the
CSM is believed to originate from the interaction of a slow wind
from the red supergiant phase with the faster wind from the blue
supergiant phase (e.g. \cite{2007Sci...315.1103M}). According to
most recent observations (\cite{2016ApJ...829...40F}), currently
the blast wave from the explosion has left the inner equatorial
ring.

\begin{figure}[t]
\begin{center}
 \includegraphics[width=13cm]{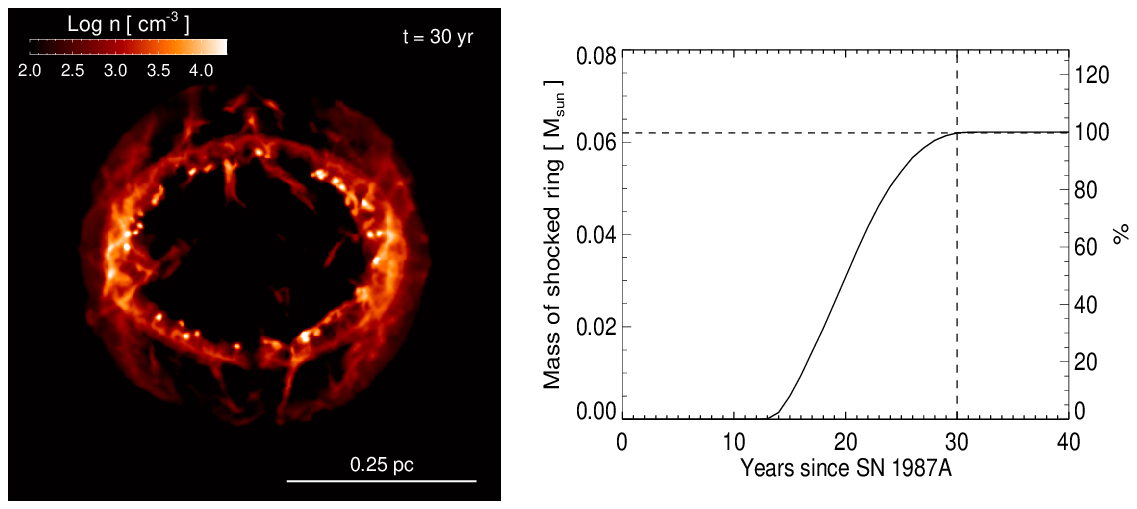}
 \caption{Left panel: three-dimensional volume rendering of the
 particle density of the shocked plasma during the interaction of
 the blast wave with the nebula, 30 years after the SN event (from
 the best-fit model of \cite{2015ApJ...810..168O}). Right panel:
 mass of the shocked ring vs. time; the vertical dashed line marks
 30 years after the SN event; the horizontal dashed line indicates
 that the whole ring has been shocked.}
   \label{hydro_30_87A}
\end{center}
\end{figure}

In a previous work, we have investigated the effects of the CSM on the
morphology of the remnant of SN\,1987A and the imprint of the SN
on the X-ray emission of its remnant through hydrodynamic simulations
(\cite{2015ApJ...810..168O}). The strategy is again that described
in Sect.~\ref{sec:strategy}. The model describes the complex structure
of the CSM accounting, in particular, for the H\,II region and the
dense equatorial ring. We have explored the space of model parameters,
searching for those best fitting the morphology, light curves, and
spectra observed in the X-ray band. We have found that a model with
explosion energy in the range $1.2-1.4\times 10^{51}$~erg and
envelope mass in the range $15-17 M_{\odot}$ is able to reproduce
the observed bolometric lightcurve of SN\,1987A during the first
250 days after the SN event. The left panel of Fig.~\ref{hydro_30_87A}
shows the 3D volume rendering of the particle density of the shocked
plasma at $t=30$~yr for the best-fit model in Orlando et al. (2015).
This model predicts a total mass of the ring $M_{\rm rg} =
0.062\,M_{\odot}$ of which $\sim 64$\% is plasma with density $n
\approx 1000$~cm$^{-3}$ and $\sim 36$\% is plasma with $n \approx
2.5\times 10^4$~cm$^{-3}$. Interestingly, these values are in very
nice agreement with those derived for the density structure of the
ring from the analysis of optical observations
(\cite{2010ApJ...717.1140M}). The model also predicts that the blast
wave has recently left the dense ring and is now propagating in a
less dense environment (see right panel in Fig.~\ref{hydro_30_87A}).
Again this is in excellent agreement with the recent findings from
the analysis of X-ray observations (\cite{2016ApJ...829...40F}) and,
also, with the fading of the ring and implied destruction inferred
from optical observations (\cite{2015ApJ...806L..19F}).

\begin{figure}[t]
\begin{center}
 \includegraphics[width=10cm]{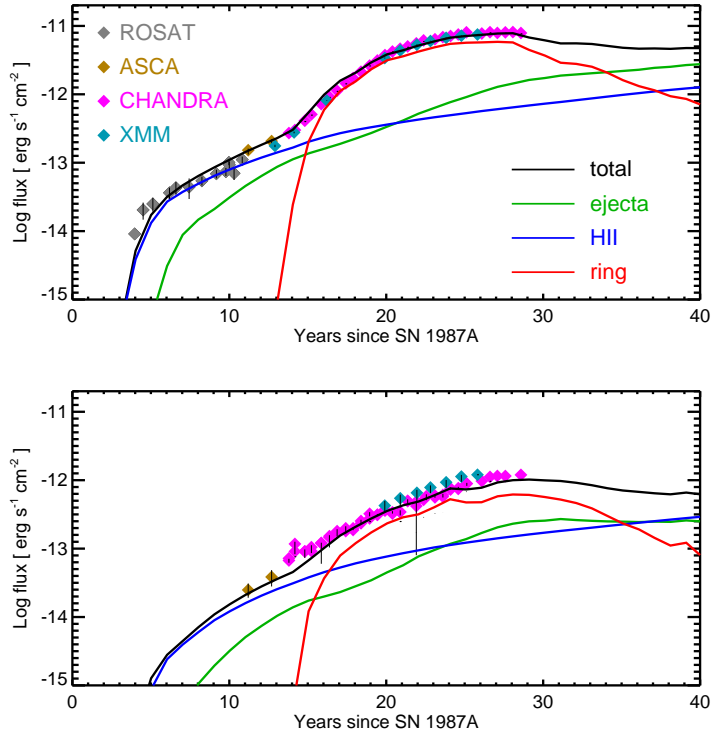}
 \caption{Observed and modeled X-ray light curves in the soft ($[0.5,
 2.0]$~keV; upper panel) and hard ($[3.0, 10]$~keV; lower panel)
 bands (adapted from Orlando et al.  2015). Green, blue, and red
 lines mark the contribution to emission from the shocked ejecta,
 the shocked plasma from the H II region, and the shocked plasma
 from the ring, respectively.}
   \label{fig_lc_87a}
\end{center} \end{figure}

From the model results, we have synthesized the X-ray emission,
deriving light curves, images, and spectra to be compared with actual
observations. Figure~\ref{fig_lc_87a} shows the light curves derived
from our best-fit model and the comparison with those observed. We
have found that the same model fitting the bolometric lightcurve
of the SN is also able to fit both the soft ($[0.5, 2.0]$~keV) and
hard ($[3.0, 10]$~keV) X-ray light curves as derived from multi-epoch
observations with ROSAT, ASCA, Chandra, and XMM-Newton X-ray
telescopes. In particular, the predictions of the model fit the
most recent Chandra observations (\cite{2016ApJ...829...40F}),
supporting the idea that the blast wave has now left the dense ring.
This study has demonstrated, for the first time, the consistency
between a physical model reproducing the observables of the SN
(namely the cause) and that reproducing the observables of the
resulting remnant (namely the effect).

\begin{figure}[t]
\begin{center}
 \includegraphics[width=10.5cm]{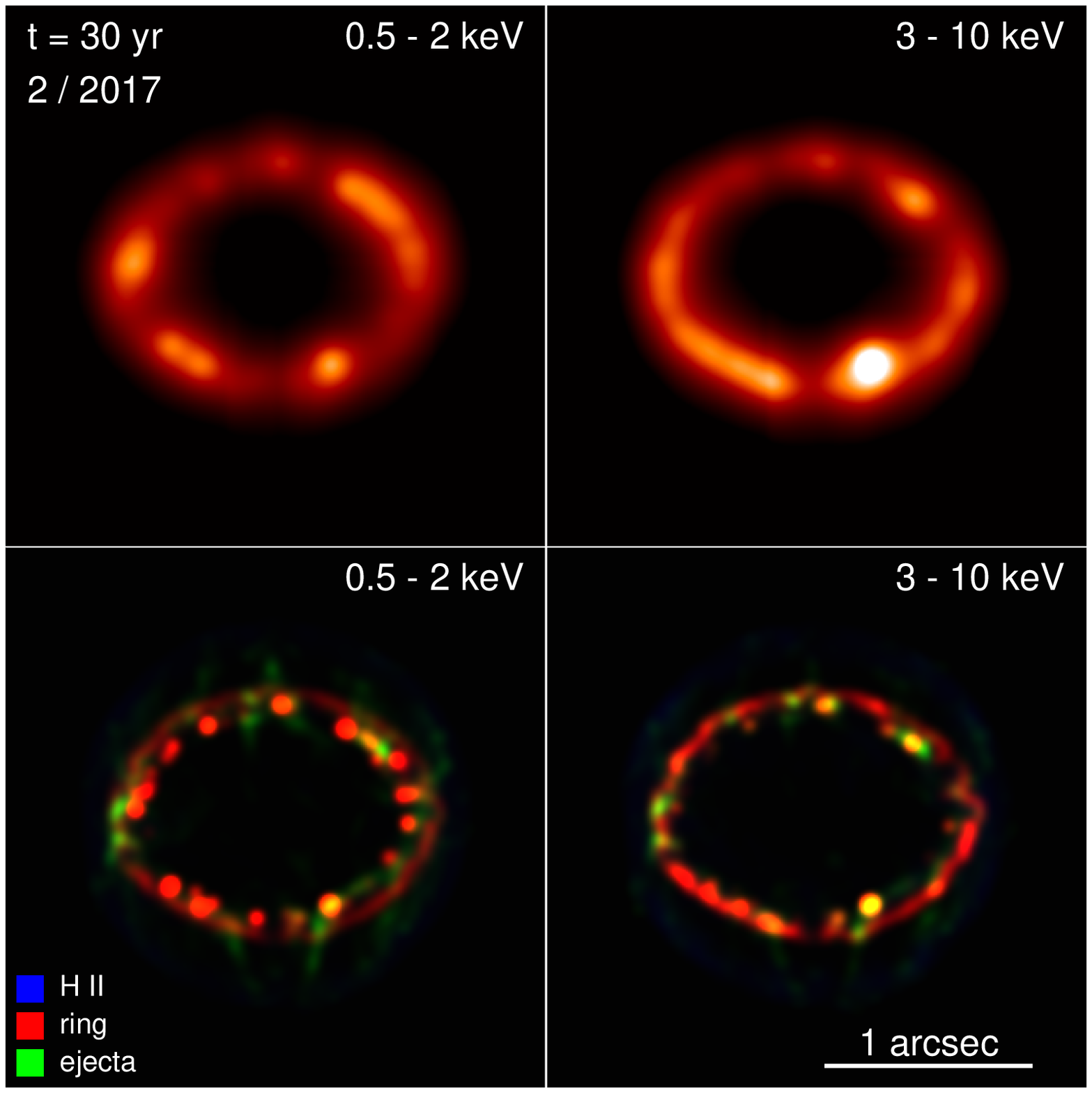}
 \caption{Synthetic maps of X-ray emission in the soft ($[0.5,
 2.0]$~keV; on the left) and hard ($[3.0, 10]$~keV; on the right)
 bands integrated along the line of sight, 30 years after the SN
 event (from the best-fit model of \cite{2015ApJ...810..168O});
 each image has been normalized to its maximum for visibility. In
 the upper panels the images have been convolved with a Gaussian
 of size 0.15 arcsec to approximate the spatial resolution of Chandra
 observations. In the lower panels, the figure shows three-color
 composite images of the X-ray emission smoothed with a Gaussian
 of size 0.025 arcsec; the colors in the composite show the
 contribution to emission from the different shocked plasma components,
 namely the ejecta (green), the ring (red), and the H II region
 (blue).}
   \label{emiss_30_87A}
\end{center} \end{figure}

The model includes also some passive tracers to follow the evolution
of the different plasma components (namely ejecta, H\,II region,
and dense ring) and their contribution to X-ray emission (see
Figs.~\ref{fig_lc_87a} and \ref{emiss_30_87A}). We have identified
three main phases in the evolution of the remnant of SN\,1987A. The
first starts when the ejecta reach the H\,II region and last for
about 13 years. In this phase the emission is dominated by the
shocked H\,II region with a smaller contribution from shocked ejecta.
Then the blast wave starts to propagate through the dense ring
($\approx 13$ yr after the SN event) and the X-ray emission becomes
largely dominated by the shocked plasma from the ring (see
Fig.~\ref{emiss_30_87A}). In this phase the contribution from shocked
ejecta continue to increase faster than that of shocked plasma from
the H\,II region, but still much lower than the contribution from
shocked plasma from the ring (see Fig.~\ref{fig_lc_87a}). The model
predicts that the blast wave has left the disk in the last few years
and is now propagating through a less dense medium. As a consequence,
the contribution from shocked ring will shortly decrease while the
contribution from shocked ejecta will gradually increase becoming
the dominant component in about $4-5$ years (see Fig.~\ref{fig_lc_87a}).
During the first phase (H\,II region dominated), we have found that
the shape of the X-ray lightcurve reflects the structure of the
outer ejecta thus identifying the imprint of SN\,1987A on the X-ray
emission of its remnant. In the second phase (ring dominated), the
comparison between model results and observations allowed us to
constrain the structure and geometry of the CSM (see
Fig.~\ref{emiss_30_87A}). In the third phase (ejecta dominated),
expected in the next future, our model predicts that we will be
able to investigate the structure and chemical stratification of
the ejecta.

\section{Conclusions}

There is a growing interest in Astrophysics on studies aimed at
linking the morphological properties of SNRs to their progenitor SNe.
In fact SNRs morphology and properties are expected to reflect the
physical and chemical properties of the progenitor SNe and of the
environment in which blast waves travel. Multi-wavelength/multi-messenger
observations of SNRs, therefore, should encode information about
the physical and chemical properties of both the stellar debris and
the surrounding CSM. In the first case, we could have the possibility
to investigate the energetics and dynamics of the SN progenitor and
the anisotropies developed during the complex phases of the SN
explosion; in the latter we could derive important clues on the
final stages of stellar evolution.  Linking SNe to SNRs, therefore,
has breakthrough potential to open new exploring windows on SN and
SNR issues, especially in view of the spatially resolved high-resolution
spectroscopy capability of the forthcoming Athena satellite
(\cite{2013arXiv1306.2307N}).

In this contribution we have reviewed our recent achievements in the
study of the SN-SNR connection. In particular, we have devised a
strategy based on 3D hydrodynamic models to link the morphological
properties observed in SNRs to the structure of the ejecta soon
after the progenitor SN explosions and to the structure of the
environment in which the remnant expands. In the case of Cas\,A,
we have shown that our approach was very effective in deriving the
physical properties of the main anisotropies developed in the
aftermath of the SN explosion and responsible for the observed
non-uniform distribution of ejecta. In the case of SN\,1987A, we
have demonstrated that the physical model reproducing the main
observables of the SN reproduces also the X-ray emission of the
subsequent expanding remnant, thus bridging the gap between SNe and
SNRs. The study was very effective also to identify the imprint of
SN\,1987A on the X-ray emission of its remnant and to constrain the
structure and geometry of the surrounding nebula, formed in the
latest stage of the star progenitor's evolution.

Our study has shown that detailed and complete numerical models
accounting for the 3D nature of the phenomenon are necessary to
successfully decipher the observations and, therefore, to connect
the observed morphology of SNRs to the physics of the progenitor
SN engine. However, deciphering observations might depend critically
on the models.  The latter should connect the stellar progenitor,
the SN, and the SNR self-consistently. The models should be guided
by observational facts and they should account for the dynamics,
energetics, and spectral properties of SNe and SNRs. Finally it is
highly desirable to strongly improve the synergy and communication
between the SN and SNR communities. The IAU Symposium 331, bringing
together theorists and observers from these communities, is an
excellent example of how to accomplish this task.

\acknowledgments
The authors acknowledge partial support from the IAU to participate
to the IAU Symposium 331. This paper was partially funded by the
PRIN INAF 2014 grant ``Filling the gap between supernova explosions
and their remnants through magnetohydrodynamic modeling and high
performance computing''. The software used in this work was, in
part, developed by the U.S.  Department of Energy-supported Advanced
Simulation and Computing/Alliance Center for Astrophysical Thermonuclear
Flashes at the University of Chicago. We acknowledge that the results
of this research have been achieved using the PRACE Research
Infrastructure resource MareNostrum III based in Spain at the
Barcelona Supercomputing Center (PRACE Award N.2012060993) and the
FERMI cluster based in Italy at CINECA (ISCRA Award N. HP10BI36DG,2012).

\end{document}